\documentclass[10pt]{article}
\usepackage{graphicx}
\usepackage{amsmath}
\usepackage{amssymb}
\usepackage{caption2}
\begin{document}
\bibliographystyle{prsty}
\begin{center}
{\large {\bf \sc{  Analysis of  the heavy quarkonium states $h_c$ and $h_b$  with  QCD sum rules }}} \\[2mm]
Zhi-Gang Wang \footnote{E-mail,zgwang@aliyun.com.  }     \\
 Department of Physics, North China Electric Power University,
Baoding 071003, P. R. China
\end{center}

\begin{abstract}
In this article, we take the tensor currents $\bar{Q}(x)\sigma_{\mu\nu}Q(x)$ to interpolate the $P$-wave spin-singlet heavy quarkonium states $h_Q$, and study the masses and decay constants with  the Borel sum rules and moments sum rules.
The  masses and decay constants from the Borel sum rules and moments sum rules are consistent with each other, the masses are also consistent with the experimental data. We can take the decay constants as   basic input parameters and study  other phenomenological quantities   with the three-point correlation functions via the QCD sum rules.  The heavy quarkonium states $h_Q$ couple potentially to the tensor currents $\bar{Q}(x)\sigma_{\mu\nu}Q(x)$,  and have the quark structure $\epsilon^{ijk}\xi^{\dagger}\sigma^k\zeta$ besides the quark structure $ik_2^i \xi^{\dagger}\sigma \cdot (\vec{k}_1-\vec{k}_2)\zeta $.
In calculations, we take into account the leading-order, next-to-leading-order perturbative contributions, and the gluon condensate, four-quark condensate contributions in the operator product expansion. The analytical expressions of the perturbative QCD spectral densities have applications in studying the two-body decays of a boson to two fermions with the vertexes $\sigma_{\mu\nu}\gamma_5$ and $\sigma_{\mu\nu}$.
\end{abstract}



\section{Introduction}

In 2011,  the BABAR collaboration  observed  evidences  for the spin-singlet bottomonium state $h_b(1\rm{P})$  in the sequential decays $\Upsilon(3{\rm S}) \to \pi^0 h_b(1{\rm P})$, $h_b(1{\rm P}) \to \gamma \eta_b(1{\rm S})$ \cite{Babar1102}.
Later, the Belle collaboration   reported the first observation of the spin-singlet bottomonium states $h_b(1\rm{P})$ and $h_b(2\rm{P})$ with the significances
of $5.5\,\sigma$ and $11.2\,\sigma$ respectively  in the collisions
$e^+e^- \to h_b(n\rm{P})\pi^+\pi^-$  at energies near the $\Upsilon(5{\rm S})$ resonance, and  determined the masses
$M_{h_b(1\rm{P})}=\left(9898.3\pm1.1{}^{+1.0}_{-1.1}\right) \,\rm{MeV}$ and $M_{h_b(2\rm{P})}=\left(10259.8\pm0.6{}^{+1.4}_{-1.0}\right) \, \rm{MeV}$ \cite{Belle1103}.   On the other hand, the mass of the spin-singlet charmonium state $h_c(1\rm{P})$ has  been updated from time to time since its first observation in the $p\bar{p}$ collisions by the  R704  collaboration \cite{Hc1986},  the average value listed in the  Review of Particle Physics is $M_{h_c(1\rm{P})}=(3525.41\pm0.16)\,\rm{MeV}$ \cite{PDG}.

The  heavy quarkonium states  play an important role both in studying the interplays between the perturbative and nonperturbative QCD
  and in understanding the heavy quark dynamics due to  absence  of the light quark contaminations. In this article, we study the heavy quarkonium states $h_c$ and $h_b$ with  the QCD sum rules, explore  their quark structures,
and make predictions for the masses to be confronted  with experimental data.
The QCD sum rules is a powerful (nonperturbative) theoretical tool  in
 studying   the heavy quarkonium states \cite{SVZ79,Reinders85},   the existing works focus on the $S$-wave heavy quarkonium states $J/\psi$, $\eta_c$, $\Upsilon$, $\eta_b$, and the $P$-wave spin-triplet heavy quarkonium states $\chi_{cJ}$, $\chi_{bJ}$, $J=0,1,2$, while the works on the  $P$-wave spin-singlet heavy quarkonium states $h_c$ and $h_b$  are few \cite{Reinders85,NarisonBook}. On the other hand, the heavy quarkonium spectrum have been studied extensively  by the (potential) nonrelativistic QCD, and the existing works also focus on the $S$-wave heavy quarkonium states and  $P$-wave spin-triplet heavy quarkonium states \cite{NRQCD},
the works on the  $P$-wave spin-singlet heavy quarkonium states $h_Q$  are  few \cite{NRQCD-hQ}. In the (potential) nonrelativistic QCD, the fine splittings  and hyperfine splittings among the  heavy quarkonium states are treated  perturbatively.

The tensor currents $\bar{Q}(x)\sigma_{\mu\nu}Q(x)$ and axialvector currents $\bar{Q}(x)\gamma^{\mu}\gamma_5Q(x)$ without derivatives have the following properties under the parity and charge-conjunction transforms,
 \begin{eqnarray}
\bar{Q}(x)\sigma^{\mu\nu}Q(x)&\stackrel{P}{\longrightarrow}&\bar{Q}(\tilde{x})\sigma_{\mu\nu}Q(\tilde{x}) \, ,  \nonumber \\
\bar{Q}(x)\sigma^{\mu\nu}Q(x)&\stackrel{C}{\longrightarrow}&-\bar{Q}(x)\sigma^{\mu\nu}Q(x) \, , \nonumber \\
\bar{Q}(x)\gamma^{\mu}\gamma_5Q(x)&\stackrel{P}{\longrightarrow}&-\bar{Q}(\tilde{x})\gamma_{\mu}\gamma_5Q(\tilde{x}) \, ,  \nonumber \\
\bar{Q}(x)\gamma^{\mu}\gamma_5Q(x)&\stackrel{C}{\longrightarrow}&\bar{Q}(x)\gamma^{\mu}\gamma_5Q(x) \, .
\end{eqnarray}
where $x^\mu=(t,\vec{x})$ and $\tilde{x}^\mu=(t,-\vec{x})$.
The $P$-wave spin-singlet heavy quarkonium states $h_Q$ have the spin-parity-charge-conjunction $J^{PC}=1^{+-}$,
the axialvector currents $\bar{Q}(x)\gamma^{\mu}\gamma_5Q(x)$ couple potentially   to the axialvector heavy quarkonium states $\chi_{c1}$ and $\chi_{b1}$, which have the quantum numbers $J^{PC}=1^{++}$ rather than $1^{+-}$, the tensor currents are superior to the axialvector currents in studying  the $h_Q$. In Ref.\cite{Reinders85},
Reinders,   Rubinstein and   Yazaki study the $h_Q$ using the interpolating currents $\bar{Q}(x)\partial_\mu \gamma_5 Q(x)$ with derivatives,  and obtain the prediction $M_{h_c(1\rm{P})}=(3.51\pm0.01)\,\rm{GeV}$.

In the nonrelativistic limit, the interpolating currents are reduced to the following form,
\begin{eqnarray}
\bar{Q}\sigma^{\mu\nu}Q &\rightarrow& 2m_Q\epsilon^{ijk}\xi^{\dagger}\sigma^k\zeta  \cdots \cdots \cdots \cdots \cdot\cdot J^{PC}=1^{+-}\, ,  \nonumber\\
\bar{Q}\gamma^\mu\gamma_5 Q &\rightarrow&  2m_Q\xi^{\dagger}\sigma^i\zeta \cdots \cdots\cdots \cdots \cdots\cdots J^{PC}=1^{++}\, ,\nonumber\\
\bar{Q}\gamma_5\partial^\mu Q &\rightarrow& ik_2^i \xi^{\dagger}\sigma \cdot (\vec{k}_1-\vec{k}_2)\zeta  \cdots \cdot \cdots \cdots J^{PC}=1^{+-}\, ,
\end{eqnarray}
where the $\xi$ and $\zeta$ are the two-component spinors of the heavy quark fields $\bar{Q}$ and $Q$ respectively, the $\vec{k}_1$ and $\vec{k}_2$ are the three-vectors of the  heavy quark fields $\bar{Q}$ and $Q$ respectively, and the $\sigma^i$ are the pauli matrixes. From Eq.(2), we can see that the interpolating currents $\bar{Q}\sigma^{\mu\nu}Q$ and $\bar{Q}\gamma_5\partial^\mu Q$ both have the correct quantum numbers of the heavy quarkonium states $h_Q$, therefor  they both couple potentially to the $h_Q$. It is interesting to study whether or not the $h_Q$ have the quark structure $\epsilon^{ijk}\xi^{\dagger}\sigma^k\zeta$ besides the quark structure $ik_2^i \xi^{\dagger}\sigma \cdot (\vec{k}_1-\vec{k}_2)\zeta$.

In the QCD sum rules, additional partial  derivative $\partial^\mu$ in the interpolating currents lead  to additional power of $s$ in the spectral densities $\rho(s)$ of the two-point correlation functions,  which   enhances the continuum contributions even if the Borel depression is  taken into account, see Fig.1. In the limit $m_Q\rightarrow 0$, the spectral densities $\rho(s)$ are of the orders $\mathcal{O}(1)$ and $\mathcal{O}(s^2)$ for the currents $\bar{Q}\sigma^{\mu\nu}Q$ and $\bar{Q}\gamma_5\partial^\mu Q$, respectively, and we prefer  constructing quark  currents without partial  derivatives.
In this article, we interpolate  the singlet heavy quarkonium states $h_Q$ with the tensor currents $\bar{Q}\sigma^{\mu\nu}Q$, calculate the masses and decay constants (or pole residues). The decay constants are  basic input parameters in studying the $h_c DD^*$, $ h_c D_sD_s^*$, $h_c D^*D^*$, $h_c D_s^*D_s^*$ vertexes and $h_c \to D,\, D_s,\,D^*,\,D_s^*$ form-factors  with three-point correlation functions using the QCD sum rules,

\begin{figure}
 \centering
 \includegraphics[totalheight=5cm,width=8cm]{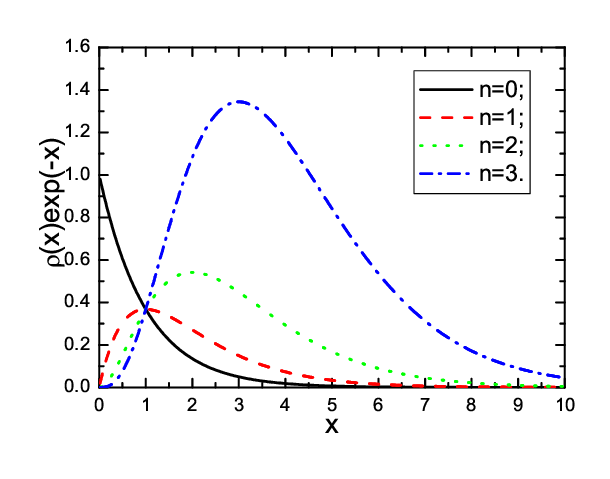}
    \caption{ The Borel parameter depressed density $\rho(x)\exp(-x)=x^n\exp(-x)$ with $x=\frac{s}{T^2}$, where the $T^2$ is the Borel parameter and the $\rho(x)$ is the  spectral density. }
\end{figure}

The article is arranged as follows:  we derive the QCD sum rules for
the masses and decay constants of the heavy quarkonium states $h_Q$  in Sect.2;
in Sect.3, we present the numerical results and discussions; and Sect.4 is reserved for our
conclusions.

\section{QCD sum rules for  the heavy quarkonium states   $h_Q$ }
In the following, we write down  the two-point correlation functions
$\Pi_{\mu\nu\alpha\beta}(p)$  in the QCD sum rules,
\begin{eqnarray}
\Pi_{\mu\nu\alpha\beta}(p)&=&i\int d^4x e^{ip \cdot x} \langle
0|T\left\{J_{\mu\nu}(x)J_{\alpha\beta}^{\dagger}(0)\right\}|0\rangle \, , \\
J^1_{\mu\nu}(x)&=&\bar{Q}(x)\sigma_{\mu\nu}\gamma_5 Q(x)  \, ,\nonumber\\
J^2_{\mu\nu}(x)&=&\bar{Q}(x)\sigma_{\mu\nu} Q(x)  \, ,
\end{eqnarray}
where $J_{\mu\nu}(x)=J^1_{\mu\nu}(x),J^2_{\mu\nu}(x)$, the two interpolating currents are related with each other through  the relation $\sigma^{\mu\nu}\gamma_5=\frac{i}{2}\epsilon^{\mu\nu\alpha\beta}\sigma_{\alpha\beta}$.
We decompose the correlation functions  $\Pi_{\mu\nu\alpha\beta}(p)$ as
\begin{eqnarray}
\Pi_{\mu\nu\alpha\beta}(p) &=&\Pi(p)\left(\widetilde{g}_{\mu\alpha} p_\nu p_\beta +\widetilde{g}_{\nu\beta} p_\mu p_\alpha -\widetilde{g}_{\mu\beta}p_\nu p_\alpha -\widetilde{g}_{\nu\alpha}p_\mu p_\beta\right)+\widetilde{\Pi}(p)\left(g_{\mu\alpha}g_{\nu\beta}-g_{\mu\beta}g_{\nu\alpha}\right) \, ,
\end{eqnarray}
according to Lorentz covariance,
where
\begin{eqnarray}
\widetilde{g}_{\mu\nu}&=&-g_{\mu\nu}+\frac{p_{\mu}p_{\nu}}{p^2}\, .
\end{eqnarray}
 Then we project the components $\Pi(p)$ and $\widetilde{\Pi}(p)$,
\begin{eqnarray}
\Pi(p)&=&\frac{1}{2(1-D)p^2}\left( g_{\mu\alpha}g_{\nu\beta}-\frac{D}{D-2}\widetilde{g}_{\mu\alpha}\widetilde{g}_{\nu\beta}\right)\Pi_{\mu\nu\alpha\beta}(p)\, , \nonumber\\
\widetilde{\Pi}(p)&=& \frac{1}{(D-1)(D-2)}\widetilde{g}_{\mu\alpha}\widetilde{g}_{\nu\beta}\Pi_{\mu\nu\alpha\beta}(p)\, ,
\end{eqnarray}
where $D$ is the spacetime dimension.

We can insert  a complete set of intermediate hadronic states with
the same quantum numbers as the current operators $J_{\mu\nu}(x)$ into the
correlation functions $\Pi_{\mu\nu\alpha\beta}(p)$  to obtain the hadronic representation
\cite{SVZ79,Reinders85}. After isolating the ground state
contribution from the  heavy quarkonium states   $h_Q$ , we get the following result,
\begin{eqnarray}
\Pi_{\mu\nu\alpha\beta}(p)&=&\frac{f_{h_Q}^2}{M_{h_Q}^2-p^2}\left(\widetilde{g}_{\mu\alpha} p_\nu p_\beta +\widetilde{g}_{\nu\beta} p_\mu p_\alpha -\widetilde{g}_{\mu\beta}p_\nu p_\alpha -\widetilde{g}_{\nu\alpha}p_\mu p_\beta\right) +\cdots \,  , \nonumber \\
&=&\Pi(p)\left(\widetilde{g}_{\mu\alpha} p_\nu p_\beta +\widetilde{g}_{\nu\beta} p_\mu p_\alpha -\widetilde{g}_{\mu\beta}p_\nu p_\alpha -\widetilde{g}_{\nu\alpha}p_\mu p_\beta\right) +\cdots \, \, ,
\end{eqnarray}
where the  decay constants $f_{h_Q}$ are defined by
\begin{eqnarray}
\langle 0|J^1_{\mu\nu}(0)|h_Q(p)\rangle&=&f_{h_Q}(\varepsilon_{\mu} p_{\nu}-\varepsilon_{\nu} p_{\mu}) \, , \nonumber\\
\langle 0|J^2_{\mu\nu}(0)|h_Q(p)\rangle&=&if_{h_Q}\epsilon_{\mu\nu\lambda\tau}\varepsilon^{\lambda} p^{\tau} \, ,
\end{eqnarray}
and the $\varepsilon_\mu$ are the  polarization vectors of the heavy quarkonium states $h_Q$. We choose the tensor structure
$ \widetilde{g}_{\mu\alpha} p_\nu p_\beta +\widetilde{g}_{\nu\beta} p_\mu p_\alpha -\widetilde{g}_{\mu\beta}p_\nu p_\alpha -\widetilde{g}_{\nu\alpha}p_\mu p_\beta $ to study the heavy quarkonium states   $h_Q$. In this article, we take  a simple
ground state plus continuum ansatz to approximate the phenomenological spectral densities. Experimentally, the first few radial excited quarkonium (or bottomonium) states are narrow and appear as resonance-like states rather than as continuum-like states. As the dominant contributions come from the perturbative terms and the gluon condensates play a minor important role, the higher resonance-like states can also be described by the perturbative terms and  attributed  to the continuum states, such a simple approximation (or ansatz) works well.

One may concerns the possible contaminations come from the $J=2$ tensor mesons. The $J=2$ tensor mesons $\chi_{Q2}$ couple potentially to the interpolating currents $\eta_{\mu\nu}(x)$,
\begin{eqnarray}
 \langle 0|\eta_{\mu\nu}(0)|\chi_{Q2}(p)\rangle&=&\lambda_{\chi}\varepsilon_{\mu\nu}\, ,
 \end{eqnarray}
 where
\begin{eqnarray}
\eta_{\mu\nu}(x)&=&\frac{i}{2}\left\{ \bar{Q}(x)\gamma_\mu
\left[\overrightarrow{D}_\nu(x)-\overleftarrow{D}_\nu(x)\right]Q(x)+\bar{Q}(x)\gamma_\nu\left[\overrightarrow{D}_\mu(x)-\overleftarrow{D}_\mu(x)\right]Q(x)\right\}\, ,
\end{eqnarray}
  $\overrightarrow{D}_\mu(x)=\overrightarrow{\partial}_\mu(x)-ig_sG_\mu(x)$,
 $\overleftarrow{D}_\mu(x)=\overleftarrow{\partial}_\mu(x)+ig_sG_\mu(x)$,
 $G_\mu=\frac{\lambda^n}{2}G_\mu^n$, the $\lambda^n$ are the Gell-Mann matrixes,  the $\varepsilon_{\mu\nu}$ are the polarization tensors of the $\chi_{Q2}$ mesons with the property,
\begin{eqnarray}
\sum_{\lambda}\varepsilon^*_{\alpha\beta}(\lambda,p)\varepsilon_{\mu\nu}(\lambda,p)
 &=&\frac{\widetilde{g}_{\alpha\mu}\widetilde{g}_{\beta\nu}+\widetilde{g}_{\alpha\nu}\widetilde{g}_{\beta\mu}}{2}-\frac{\widetilde{g}_{\alpha\beta}\widetilde{g}_{\mu\nu}}{3}\, .
 \end{eqnarray}
The   $J=2$ tensor mesons $\chi_{Q2}$ have no contaminations \cite{Aliev-Wang}.

We  carry out the Borel transforms (and the derivatives) with respect to the variable  $P^2=-p^2$ to obtain the Borel sum rules (and the moments sum rules),
and write down the following results at the phenomenological side,
\begin{eqnarray}
\Pi(T^2)&=&\frac{1}{\Gamma(n)}P^{2n}\left(-\frac{d}{dP^2}\right)^{n}\Pi(P^2)|_{P^2\rightarrow \infty, n\rightarrow \infty;\,P^2/n=T^2}\, , \nonumber \\
&=&\frac{1}{ T^2}\int_{4m_Q^2}^{s_0} ds \frac{{\rm{Im}}\Pi(s)}{\pi} e^{-\frac{s}{T^2}}=\frac{f_{h_Q}^2}{T^2} e^{-\frac{M_{h_Q}^2}{T^2}}\, ,\\
\Pi(n,\xi)&=&\frac{1}{\Gamma(n+1)}\left(-\frac{d}{dP^2}\right)^{n}\Pi(P^2)|_{P^2=4m_Q^2\xi} \, ,\nonumber \\
&=&\frac{1}{\pi}\int_{4m_Q^2}^{s_0} ds\frac{{\rm{Im}}\Pi(s)}{(s+ 4m_Q^2\xi)^{n+1}}=\frac{f_{h_Q}^2}{(M_{h_Q}^2+4m_Q^2\xi)^{n+1}}\, ,
\end{eqnarray}
where the $s_0$ are the continuum threshold parameters.

In the following, we briefly outline  the operator product
expansion for the correlation functions  $\Pi_{\mu\nu\alpha\beta}(p)$  in perturbative
QCD.  The Feynman diagram for the leading-order perturbative contribution is shown in Fig.2.
We calculate the diagram using the Cutkosky's rule to obtain the leading-order spectral densities $\rho_{0}(s)$,
\begin{eqnarray}
\rho_{0}(s)&=&\frac{{\rm{Im}}\Pi_0(s)}{\pi}=\frac{\sqrt{\lambda(s,m_Q^2,m_Q^2)}(s+2m_Q^2)}{4\pi^2s^2}\, ,
\end{eqnarray}
where $\lambda(a,b,c)=a^2+b^2+c^2-2ab-2bc-2ca$.

\begin{figure}
 \centering
 \includegraphics[totalheight=3cm,width=5cm]{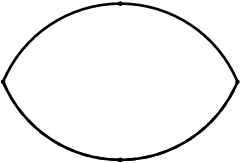}
    \caption{The leading-order perturbative contribution to the correlation function. }
\end{figure}

The Feynman diagrams for the next-to-leading-order perturbative contributions are shown in Fig.3.
Again we   calculate the diagrams using the Cutkosky's rule to obtain the   spectral densities.
There are two routines in application of  the Cutkosky's rule (or optical theorem), we resort to the routine used in Ref.\cite{Reinders85}, not the one used in Ref.\cite{Cut-2}.

\begin{figure}
 \centering
 \includegraphics[totalheight=2.7cm,width=14cm]{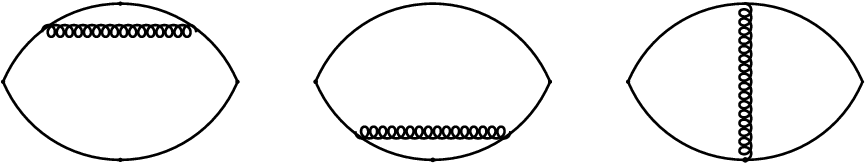}
    \caption{The next-to-leading order perturbative contributions to the correlation function. }
\end{figure}

There are ten possible cuts,  see Fig.4 and Fig.5.
The six cuts shown in Fig.4  attribute to virtual gluon emissions and correspond to the self-energy corrections and vertex corrections respectively.
 We calculate the  one-loop quark self-energy corrections directly  using the dimensional regularization and
choose the on-shell renormalization scheme to subtract the divergences so as to implement the wave-function renormalization and mass renormalization.
Then we take into account all  contributions come from the six cuts shown in Fig.4 by the following simple replacement for each vertex $\sigma_{\mu\nu} \gamma_5$ in the interpolating currents,
\begin{eqnarray}
\bar{u}(p_1)\sigma_{\mu\nu} \gamma_5 u(p_2)&\rightarrow& \bar{u}(p_1)\sigma_{\mu\nu} \gamma_5 u(p_2)+\bar{u}(p_1)\widetilde{\Gamma}_{\mu\nu} \gamma_5 u(p_2)   \nonumber \\
&=& \sqrt{Z_1}\sqrt{Z_2}\bar{u}(p_1)\sigma_{\mu\nu}\gamma_5 u(p_2)+\bar{u}(p_1)\Gamma_{\mu\nu}\gamma_5 u(p_2)-\delta Z_{\sigma}\bar{u}(p_1)\sigma_{\mu\nu}\gamma_5 u(p_2)\nonumber \\
&=& \bar{u}(p_1)\sigma_{\mu\nu}\gamma_5 u(p_2)\left(1+\frac{1}{2}\delta Z_1+\frac{1}{2}\delta Z_2 -\delta Z_{\sigma}\right)+\bar{u}(p_1)\Gamma_{\mu\nu}\gamma_5 u(p_2)\, ,
\end{eqnarray}
where
\begin{eqnarray}
Z_i&=&1+\delta Z_{i}=1+\frac{4}{3}\frac{\alpha_s}{\pi}\left(-\frac{1}{4\varepsilon_{\rm UV}}+\frac{1}{2\varepsilon_{\rm IR}}+\frac{3}{4}\log\frac{m_Q^2}{4\pi\mu^2}+\frac{3}{4}\gamma-1\right)\, ,
\end{eqnarray}
are the wave-function renormalization constants come from the self-energy corrections, see Fig.6;
\begin{eqnarray}
\Gamma_{\mu\nu} &=& \frac{4}{3}g_s^2\int_0^1 dx \int_0^{1-x}dy \int \frac{d^D k_E}{(2\pi)^D}  \frac{2\Gamma(3)}{\left[k^2_E+(x p_1+y p_2)^2\right]^3} \nonumber\\
&&\left\{\left[s-2m_Q^2-(x+y)(s-4m_Q^2)\right]\sigma_{\mu\nu} -(x+y)m_Qi \left[ (p_1+p_2)_\mu\gamma_\nu-(p_1+p_2)_\nu\gamma_\mu\right] \right\}\, ,\nonumber\\
\end{eqnarray}
comes from the vertex corrections, see Fig.7; the counterterm $\delta Z_\sigma$ comes from  renormalization of the operator $\bar{Q}\sigma_{\mu\nu}\gamma_5Q$,
\begin{eqnarray}
Z_1^{-\frac{1}{2}}Z_2^{-\frac{1}{2}}\left(\bar{Q}\sigma_{\mu\nu}\gamma_5Q\right)_0&=&Z_{\sigma}\left(\bar{Q}\sigma_{\mu\nu}\gamma_5Q\right)_r=\left(1+\delta Z_{\sigma}\right)\left(\bar{Q}\sigma_{\mu\nu}\gamma_5Q\right)_r \, ,
\end{eqnarray}
where the subindex $0$ denotes the bare quantity and the $r$ denotes the renormalized quantity.
 Here $\gamma$ is the Euler constant, $\mu^2$ is the energy scale, and the Euclidean momentum $k_E=(k_1,k_2,k_3,k_4)$.
 In this article, we take the dimension  $D=4-2\varepsilon_{\rm UV}=4+2\varepsilon_{\rm IR}$ to regularize the ultraviolet and infrared divergences respectively, and add the energy scale factors $\mu^{ 2\varepsilon_{\rm UV}}$ or $\mu^{ -2\varepsilon_{\rm IR}}$ when necessary.

We carry out the integral over the variables $x$, $y$ and $k_E$ to obtain
\begin{eqnarray}
\widetilde{\Gamma}_{\mu\nu}\gamma_5 &=& \frac{1}{3}\frac{\alpha_s}{\pi}\sigma_{\mu\nu}\gamma_5 f(s)+\frac{1}{3}\frac{\alpha_s}{\pi}i \left[ (p_1+p_2)_\mu\gamma_\nu-(p_1+p_2)_\nu\gamma_\mu\right]\gamma_5\frac{4m_Q}{\sqrt{\lambda(s,m_Q^2,m_Q^2)}} \log\left(\frac{1+\omega}{1-\omega}\right) \nonumber\\
&&-\delta Z_{\sigma} \sigma_{\mu\nu}\gamma_5\, ,
\end{eqnarray}
where
\begin{eqnarray}
f(s) &=& \overline{f}(s)- \frac{1}{\varepsilon_{\rm UV}}+ \frac{2}{\varepsilon_{\rm IR}} +3\log\frac{m_Q^2}{4\pi\mu^2}+3\gamma-4-\frac{2(s-2m_Q^2)}{\sqrt{\lambda(s,m_Q^2,m_Q^2)}}\log\left(\frac{1+\omega}{1-\omega}\right)  \nonumber\\
&& \left(\frac{1}{\varepsilon_{\rm IR}} +\log\frac{s}{4\pi\mu^2}+\gamma\right) \, ,\nonumber\\
\overline{f}(s)&=& \frac{2(s-2m_Q^2)}{\sqrt{\lambda(s,m_Q^2,m_Q^2)}}\left\{\frac{1}{2}\log^2(1-\omega^2)-2\log^2(1+\omega)+2\log2\log\left(\frac{1+\omega}{1-\omega}\right)\right.\nonumber\\
&&\left.-2{\rm Li_2}\left( \frac{2\omega}{1+\omega}\right)+\pi^2\right\}+\frac{4(s-4m_Q^2)}{\sqrt{\lambda(s,m_Q^2,m_Q^2)}}\log\left(\frac{1+\omega}{1-\omega}\right) \, ,
\end{eqnarray}
 $s=p^2$, $\omega=\sqrt{1-\frac{4m_Q^2}{s}}$, and ${\rm Li_2(x)}=-\int_0^x dt \frac{\log(1-t)}{t}$.

\begin{figure}
\centering
 \includegraphics[totalheight=3.7cm,width=14cm]{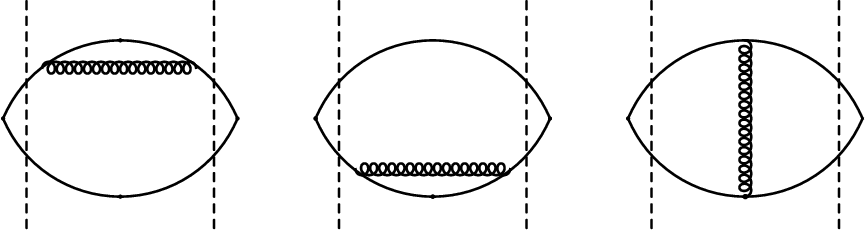}
    \caption{Six possible cuts correspond to virtual gluon emissions. }
\end{figure}

\begin{figure}
 \centering
 \includegraphics[totalheight=3.7cm,width=14cm]{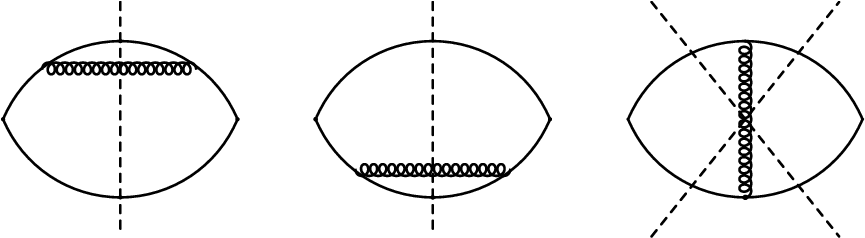}
    \caption{Four possible cuts correspond to  real gluon emissions. }
\end{figure}

\begin{figure}
 \centering
 \includegraphics[totalheight=2cm,width=8cm]{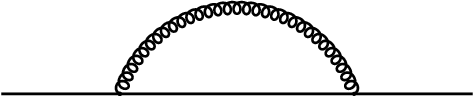}
    \caption{The quark self-energy correction. }
\end{figure}

\begin{figure}
 \centering
 \includegraphics[totalheight=3cm,width=4cm]{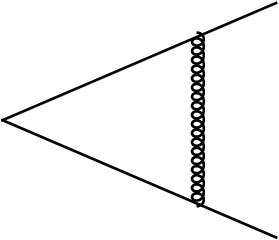}
    \caption{The vertex correction. }
\end{figure}

\begin{figure}
 \centering
 \includegraphics[totalheight=3cm,width=7cm]{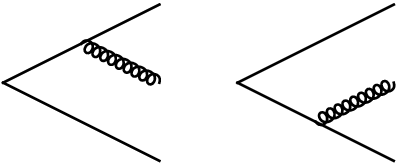}
    \caption{The amplitudes for the real gluon emissions. }
\end{figure}

The total contributions come from the virtual gluon emissions (see Fig.4) to imaginary parts of the correlation functions  can be expressed in the following form,
 \begin{eqnarray}
\frac{{\rm Im}\Pi^V_{1}(s)}{\pi}&=&\frac{4}{3}\frac{\alpha_s}{\pi}\rho_{0}(s)\left\{-\frac{1}{2\varepsilon_{\rm UV}}+\frac{1}{\varepsilon_{\rm IR}}+\frac{3}{2}\log\frac{m_Q^2}{4\pi\mu^2}+\log\frac{\lambda(s,m_Q^2,m_Q^2)}{4\pi s\mu^2}+\frac{5}{2}\gamma-4 +\frac{1}{2}\overline{f}(s)\right.\nonumber\\
&&\left.-\frac{s-2m_Q^2}{\sqrt{\lambda(s,m_Q^2,m_Q^2)}}\log\left(\frac{1+\omega}{1-\omega}\right)\left[\frac{1}{\varepsilon_{\rm IR}}+2\gamma-2+\log\frac{\lambda(s,m_Q^2,m_Q^2)}{16\pi^2\mu^4} \right] \right\} \nonumber\\
&&+\frac{4}{3}\frac{\alpha_s}{\pi}\frac{s-4m_Q^2}{12\pi^2s^2}\left\{ \sqrt{\lambda(s,m_Q^2,m_Q^2)} -(s+4m_Q^2)\log\left(\frac{1+\omega}{1-\omega}\right) \right\}-2\rho_0(s)\delta Z_\sigma \, ,
\end{eqnarray}
where
\begin{eqnarray}
\delta Z_\sigma &=&\frac{4}{3}\frac{\alpha_s}{\pi}\left(-\frac{1}{4\varepsilon_{\rm UV}}+\frac{1}{4}\log\frac{m_Q^2}{4\pi \mu^2}+\frac{\gamma}{4}\right) \, .
\end{eqnarray}

The four cuts shown in Fig.5 correspond to real gluon emissions. The scattering amplitudes for the real gluon emissions are shown explicitly in Fig.8. From Fig.8,
we can write down the scattering amplitude $T^a_{\mu\nu\alpha}(p)$,
\begin{eqnarray}
T^a_{\mu\nu\alpha}(p)&=&\bar{u}(p_1)\left\{ ig_s \frac{\lambda^a}{2}\gamma_\alpha \frac{i}{\!\not\!{p}_1+\!\not\!{k}-m_Q}\sigma_{\mu\nu}\gamma_5+\sigma_{\mu\nu}\gamma_5\frac{i}{-\!\not\!{p}_2-\!\not\!{k}-m_Q}ig_s\frac{\lambda^a}{2}\gamma_\alpha\right\}v(p_2)\,  ,
\end{eqnarray}
  then we obtain the corresponding contributions ${\rm Im}\Pi^R_1(s)$  to the imaginary parts  of the correlation functions  with
 optical theorem,
\begin{eqnarray}
\frac{{\rm Im}\Pi^R_1(s)}{\pi}&=&-\frac{1}{2\pi}  \int \frac{d^{D-1}\vec{k}}{(2\pi)^{D-1}2E_k}\frac{d^{D-1}\vec{p}_1}{(2\pi)^{D-1}2E_{p_1}}\frac{d^{D-1}\vec{p}_2}{(2\pi)^{D-1}2E_{p_2}}(2\pi)^{D}\delta^D(p-k-p_1-p_2)\nonumber\\
&&{\rm Tr}\left\{T^a_{\mu\nu\lambda}(p)T^{a\dagger}_{\alpha\beta\tau}(p)\right\}g^{\lambda\tau}\frac{1}{2(1-D)p^2}\left( g^{\mu\alpha}g^{\nu\beta}-\frac{D}{D-2}\widetilde{g}^{\mu\alpha}\widetilde{g}^{\nu\beta}\right)   \nonumber\\
&=&\frac{8g_s^2}{3\pi s}  \int \frac{d^{D-1}\vec{k}}{(2\pi)^{D-1}2E_k}\frac{d^{D-1}\vec{p}_1}{(2\pi)^{D-1}2E_{p_1}}\frac{d^{D-1}\vec{p}_2}{(2\pi)^{D-1}2E_{p_2}}(2\pi)^{D}\delta^D(p-k-p_1-p_2)\nonumber\\
&&\left\{(s+2m_Q^2)\left[\frac{s-2m_Q^2}{k\cdot p_1 k\cdot p_2}-\frac{m_Q^2}{(k\cdot p_1)^2} -\frac{m_Q^2}{(k\cdot p_2)^2}-\frac{s-K^2}{k\cdot p_1 k\cdot p_2}\right]  +\frac{(s-K^2)^2}{2k\cdot p_1 k\cdot p_2} \right.\nonumber\\
&&\left. -\frac{8(s-K^2)}{s}+\varepsilon_{\rm IR}\frac{s-4m_Q^2}{3}\left[\frac{s-2m_Q^2}{k\cdot p_1 k\cdot p_2}-\frac{m_Q^2}{(k\cdot p_1)^2}-\frac{m_Q^2}{(k\cdot p_2)^2} \right]\right\} \, ,
\end{eqnarray}
where we have used the identities $\sum u(p_1)\bar{u}(p_1)=\!\not\!{p}_1+m_Q$ and $\sum v(p_2)\bar{v}(p_2)=\!\not\!{p}_2-m_Q$ for the particle and antiparticle respectively, and take the notation $K^2=(p_1+p_2)^2$. We carry out  the integrals in Eq.(25) in $D=4+2\varepsilon_{\rm IR}$ dimension   to obtain the spectral densities,
\begin{eqnarray}
\frac{{\rm Im}\Pi^R_{1}(s)}{\pi}&=&\frac{4}{3}\frac{\alpha_s}{\pi}\rho_{0}(s)\left\{-\frac{1}{\varepsilon_{\rm IR}}-2\gamma+2-\log\frac{\lambda^3(s,m_Q^2,m_Q^2)}{16\pi^2m_Q^4s^2\mu^4} +\frac{3s}{\sqrt{\lambda(s,m_Q^2,m_Q^2)}}\log\left(\frac{1+\omega}{1-\omega}\right)\right.\nonumber\\
&&  +\frac{s-2m_Q^2}{\sqrt{\lambda(s,m_Q^2,m_Q^2)}}
\log\left(\frac{1+\omega}{1-\omega}\right) \left[\frac{1}{\varepsilon_{\rm IR}}+2\gamma-2+\log\frac{\lambda^3(s,m_Q^2,m_Q^2)}{16\pi^2m_Q^4s^2\mu^4}\right]\nonumber\\
&&\left.+(s-2m_Q^2)\overline{R}_{12}-R^1_{12}\right\}+\frac{4}{3}\frac{\alpha_s}{\pi}\frac{1}{\pi^2s^2}\left\{  \frac{1}{8}\sqrt{\lambda(s,m_Q^2,m_Q^2)}R^2_{12}-2R_0^1
    \right.\nonumber\\
&&\left.-\frac{s-4m_Q^2}{12} \left[\sqrt{\lambda(s,m_Q^2,m_Q^2)}-(s-2m_Q^2)\right]\right\} \, ,
\end{eqnarray}
the  expressions of the  $\overline{R}_{12}(s)$, $R_{12}^1(s)$, $R^2_{12}(s)$ and $R^1_0$ are  given explicitly in the appendix.

The total spectral densities $\rho_1(s)$ come from the virtual and real gluon emissions are $\rho_1(s)=\frac{{\rm Im}\Pi^V_{1}(s)}{\pi}+\frac{{\rm Im}\Pi^R_{1}(s)}{\pi}$,
\begin{eqnarray}
\rho_{1}(s)&=&\frac{4}{3}\frac{\alpha_s}{\pi}\rho_{0}(s)\left\{ \frac{1}{2}\overline{f}(s)+(s-2m_Q^2)\overline{R}_{12}(s)-R_{12}^1+\frac{3s}{\sqrt{\lambda(s,m_Q^2,m_Q^2)}}\log\left(\frac{1+\omega}{1-\omega}\right)\right. \nonumber\\
&&\left.+\frac{s-2m_Q^2}{\sqrt{\lambda(s,m_Q^2,m_Q^2)}}\log\left(\frac{1+\omega}{1-\omega}\right)
\log\frac{\lambda^2(s,m_Q^2,m_Q^2)}{m_Q^4 s^2}-\log\frac{\lambda^2(s,m_Q^2,m_Q^2)}{m_Q^6 s}-2 \right\} \nonumber\\
&&+\frac{4}{3}\frac{\alpha_s}{\pi}\frac{1}{\pi^2 s^2}\left\{ \frac{1}{8}\sqrt{\lambda(s,m_Q^2,m_Q^2)}R^2_{12}-2R^1_0 \right\} \nonumber\\
&&+\frac{4}{3}\frac{\alpha_s}{\pi}\frac{s-4m_Q^2}{12\pi^2 s^2}\left\{ s-2m_Q^2-(s+4m_Q^2)\log\left( \frac{1+\omega}{1-\omega}\right) \right\} \, ,
\end{eqnarray}
which are free of divergence. The spectral densities $\rho_{1}(s)$ have direct applications in studying  the $\mathcal{O}(\alpha_s)$ corrections for  the decays of a  boson into massive fermion-antifermion pairs with the vertexes $\sigma_{\mu\nu}\gamma_5$ and $\sigma_{\mu\nu}$,  see Figs.6-8.

In Figs.9-10, we present all the Feynman diagrams contribute to the gluon condensates and a typical Feynman diagram contributes to the four-quark condensates. We calculate those diagrams straightforwardly with help of the full quark propagator $S_{ij}(x)$,
\begin{eqnarray}
S_{ij}(x)&=&\frac{i}{(2\pi)^4}\int d^4k e^{-ik \cdot x} \left\{
\frac{\delta_{ij}}{\!\not\!{k}-m_Q}
-\frac{g_sG^n_{\alpha\beta}t^n_{ij}}{4}\frac{\sigma^{\alpha\beta}(\!\not\!{k}+m_Q)+(\!\not\!{k}+m_Q)
\sigma^{\alpha\beta}}{(k^2-m_Q^2)^2}+\frac{\delta_{ij}\langle g^2_sGG\rangle }{12}\right.\nonumber\\
&&\left. \frac{m_Qk^2+m_Q^2\!\not\!{k}}{(k^2-m_Q^2)^4}
+\frac{g_s D_\alpha G^n_{\beta\lambda}t^n_{ij}}{3}\frac{(\!\not\!{k}+m_Q)(f^{\lambda\beta\alpha}+f^{\lambda\alpha\beta}) (\!\not\!{k}+m_Q)}{(k^2-m_Q^2)^4}+\cdots\right\} \, ,\nonumber\\
f^{\lambda\alpha\beta}&=&\gamma^\lambda(\!\not\!{k}+m_Q)\gamma^\alpha(\!\not\!{k}+m_Q)\gamma^\beta\, ,
\end{eqnarray}
  $t^n=\frac{\lambda^n}{2}$,  the $i$, $j$ are color indexes, the $\langle g^2_sGG\rangle$
is the gluon condensate \cite{Reinders85}, and obtain the spectral densities $\rho_{\rm con}(s)$,
\begin{eqnarray}
\rho_{\rm con}(s)&=&-\frac{s}{12T^4}\langle\frac{\alpha_sGG}{\pi}\rangle\int_0^1dx \frac{x^3+(1-x)^3}{x(1-x)}\delta(s-\widetilde{m}_Q^2)\nonumber\\
&&-\frac{1}{12T^2}\langle\frac{\alpha_sGG}{\pi}\rangle\int_0^1dx \delta(s-\widetilde{m}_Q^2)-\frac{16\alpha_s^2\langle\bar{q}q\rangle^2}{81T^4}\int_0^1dx \left[\frac{1}{3}\left( 1+\frac{s}{T^2}\right) \right.\nonumber\\
&&\left.-\frac{1}{2x(1-x)}\left( 1+\frac{[x^2+(1-x)^2]s}{T^2}\right) \right]\delta(s-\widetilde{m}_Q^2) \, ,
\end{eqnarray}
where $\widetilde{m}_Q^2=\frac{m_Q^2}{x(1-x)}$.
We have used the equation of motion, $D^{\nu}G_{\mu\nu}^a=\sum_{q=u,d,s}g_s\bar{q}\gamma_{\mu}t^a q $,   and taken  the approximation  $\langle\bar{s}s\rangle=\langle\bar{q}q\rangle$ to obtain the contributions of the four-quark condensates.
In calculations, we observe that the contributions of the four-quark condensates are depressed by inverse  powers of the large Euclidean momentum $P^2$ (thereafter the Borel parameter $T^2$) and play  minor important roles, so we can neglect other diagrams contribute to the four-quark condensates of the order $\mathcal{O}(\alpha_s^2)$. Furthermore, we also neglect the contributions come from the three gluon condensates, as they are also depressed by inverse powers of the large Euclidean momentum $P^2$ and numerical coefficients. The old value (or the experiential value) estimated by  the instanton model is $\langle g_s^3f^{abc}G_aG_bG_c\rangle= 0.045\,\rm{GeV^6}$ \cite{ColangeloReview}, while   recent studies based on the moments sum rules  indicate
  $\langle g_s^3f^{abc}G_aG_bG_c\rangle=(8.8\pm5.5)\langle\alpha_s GG\rangle\approx 0.62\pm0.39\,\rm{GeV^6}$ \cite{Narison-gc-1105}. If we set the Borel parameters as $T^2=6\,\rm{GeV^2}$, the three-gluon condensate can be counted as $\langle g_s^3f^{abc}G_aG_bG_c\rangle/T^6=0.0002$ or $0.003\pm0.002$,  the contributions are very small.

\begin{figure}
 \centering
 \includegraphics[totalheight=2.7cm,width=14cm]{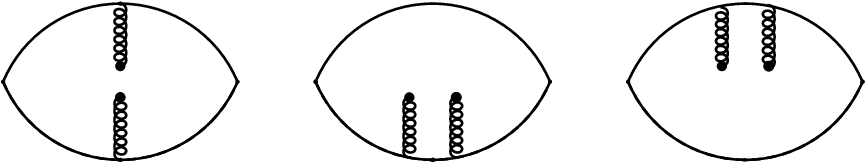}
    \caption{The diagrams contribute to the gluon condensates. }
\end{figure}
\begin{figure}
 \centering
 \includegraphics[totalheight=3cm,width=5cm]{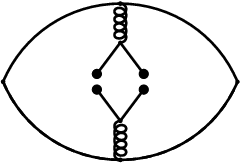}
    \caption{The typical diagram contributes to the four-quark condensate $\langle\bar{q} q\rangle^2$. }
\end{figure}

 Once analytical expressions of the QCD spectral densities are obtained,   then we can take the
quark-hadron duality and perform the Borel transforms (and the derivatives) with respect to the variable
$P^2=-p^2$ to obtain the Borel sum rules (and the moments sum rules):
\begin{eqnarray}
f_{h_Q}^2 \exp\left(-\frac{M_{h_Q}^2}{T^2}\right)&=& \int_{4m_Q^2}^{s_0} ds\left[\rho_0(s)+\rho_1(s)+\rho_{\rm con}(s) \right]\exp\left(-\frac{s}{T^2}\right) \, ,
\end{eqnarray}
\begin{eqnarray}
\frac{f_{h_Q}^2}{(M_{h_Q}^2+4m_Q^2\xi)^{n+1}}&=&\int_{4m_Q^2}^{s_0} ds \frac{\rho_0(s)+\rho_1(s)}{(s+4m_Q^2\xi)^{n+1}}-\frac{1}{12}\langle\frac{\alpha_sGG}{\pi}\rangle\int_0^1dx\frac{x^3+(1-x)^3}{x(1-x)}\nonumber\\
&&\frac{(n+1)(n+2)\widetilde{m}_Q^2}{(\widetilde{m}_Q^2+4m_Q^2\xi)^{n+3}}-\frac{1}{12}\langle\frac{\alpha_sGG}{\pi}\rangle\int_0^1dx\frac{n+1}{(\widetilde{m}_Q^2+4m_Q^2\xi)^{n+2}}
-\frac{16\alpha_s^2\langle\bar{q}q\rangle^2}{81}\nonumber\\
&&\int_0^1dx\left[\left(\frac{1}{3}-\frac{1}{2x(1-x)}\right)\frac{(n+1)(n+2)}{(\widetilde{m}_Q^2+4m_Q^2\xi)^{n+3}} +\left(\frac{1}{3}-\frac{x^2+(1-x)^2}{2x(1-x)}\right)\right.\nonumber\\
&&\left.\frac{(n+1)(n+2)(n+3)\widetilde{m}_Q^2}{(\widetilde{m}_Q^2+4m_Q^2\xi)^{n+4}} \right] \, .
\end{eqnarray}

 We can eliminate the decay constants $f_{h_Q}$ and obtain the QCD sum rules for the masses of the heavy quarkonium states $h_Q$,
 \begin{eqnarray}
 M_{h_Q}^2&=& \frac{\int_{4m_Q^2}^{s_0} ds\frac{d}{d \left(-1/T^2\right)}\left[\rho_0(s)+\rho_1(s)+\rho_{\rm con}(s)\right]\exp\left(-\frac{s}{T^2}\right)}{\int_{4m_Q^2}^{s_0} ds \left[\rho_0(s)+\rho_1(s)+\rho_{\rm con}(s)\right]\exp\left(-\frac{s}{T^2}\right)}\, , \\
 M_{h_Q}^2&=& \frac{\Pi(n-1,\xi)}{\Pi(n ,\xi)}-4m_Q^2\xi\, .
\end{eqnarray}

\section{Numerical results and discussions}
From the experimental data  $M_{h_b(1\rm{P})}=\left(9898.3\pm1.1{}^{+1.0}_{-1.1}\right)\, \rm{MeV}$, $M_{h_b(2\rm{P})}=\left(10259.8\pm0.6{}^{+1.4}_{-1.0}\right)\, \rm{MeV}$ \cite{Belle1103}, $M_{h_c(1\rm{P})}=\left(3525.41\pm0.16\right)\,\rm{MeV}$, $M_{\chi_{c2}(2\rm{P})}=\left(3927.2\pm2.6\right)\,\rm{MeV}$ \cite{PDG},
we obtain the continuum threshold parameters
$s^0_{h_c}=(16\pm1)\,\rm{GeV}^2$ and $s^0_{h_b}=(105\pm2)\,\rm{GeV}^2$ approximately.
  The quark condensate is determined by the Gell-Mann-Oakes-Renner relation, we take  the standard value $\langle
\bar{q}q \rangle=-(0.24\pm0.01\, \rm{GeV})^3$  at the energy scale  $\mu=1\, \rm{GeV}$
\cite{SVZ79,Reinders85,ColangeloReview}.
 The value of the gluon condensate $\langle \frac{\alpha_s
GG}{\pi}\rangle $ has been updated from time to time, and changes
greatly \cite{NarisonBook}, we take
    the recently updated value $\langle \frac{\alpha_s GG}{\pi}\rangle=(0.022 \pm
0.004)\,\rm{GeV}^4 $, and neglect the uncertainty.

In this article, we calculate the   perturbative $\mathcal{O}(\alpha_s)$ corrections $\rho_1(s)$ in the on-shell renormalization scheme, and take the pole masses.
The pole masses and the $\overline{MS}$ masses
have the relation  $m_Q
=\overline{m}_Q(\overline{m}_Q^2)\left[1+\frac{4 \alpha_s(\overline{m}_Q^2)}{3\pi}+\cdots\right]$.
The $\overline{MS}$  masses  have been studied extensively  by the QCD sum rules and Lattice QCD \cite{PDG,NarisonBook,ColangeloReview,Ioffe2005}. The values listed  in the Review of Particle Physics are $\overline{m}_c(\overline{m}_c^2)=1.275\pm 0.025\,\rm{GeV}$ and $\overline{m}_b(\overline{m}_b^2)=4.18 \pm0.03\,\rm{GeV}$ \cite{PDG},
 which correspond to the pole masses $m_c=(1.67 \pm 0.07)\, \rm GeV$ and  $m_b=(4.78 \pm 0.06)\, \rm GeV$.  The recent studies based on the QCD sum rules  \cite{Narison-gc-1105,Chetyrkin-2009}, the
 nonrelativistic large-n $\Upsilon$ sum rules with renormalization group improvement \cite{Hoang-2012}  and the lattice QCD \cite{McNeile-latt} indicate (slightly) different values. In this article, we choose the values $m_b=4.80\, \rm{GeV}$ and $m_c=1.55\,\rm{GeV}$, the  uncertainties will be discussed later. Furthermore, we set the energy scale to be $\mu=m_c$ and $m_b$ for the heavy quarkonium states $h_c$ and $h_b$, respectively, and take the $\alpha_s(\mu)$ from the Particle Data Group,
 \begin{eqnarray}
\alpha_s(\mu)&=&\frac{1}{b_0t}\left[1-\frac{b_1}{b_0^2}\frac{\log t}{t} +\frac{b_1^2(\log^2{t}-\log{t}-1)+b_0b_2}{b_0^4t^2}\right]\, ,
\end{eqnarray}
  where $t=\log \frac{\mu^2}{\Lambda^2}$, $b_0=\frac{33-2n_f}{12\pi}$, $b_1=\frac{153-19n_f}{24\pi^2}$, $b_2=\frac{2857-\frac{5033}{9}n_f+\frac{325}{27}n_f^2}{128\pi^3}$,  $\Lambda=213\,\rm{MeV}$, $296\,\rm{MeV}$  and  $339\,\rm{MeV}$ for the flavors  $n_f=5$, $4$ and $3$, respectively  \cite{PDG}.

If we take the Borel parameters as $T^2=(5.5-6.5)\,\rm{GeV}^2$ and $(11-13)\,\rm{GeV}^2$ in the channels $h_c$ and $h_b$, respectively,   the pole contributions are about $(51-69)\%$ and $(50-69)\%$, respectively, see Fig.11,  it is reliable to extract the ground state  masses. In Fig.11, we plot the pole contributions with variations of the Borel parameters $T^2$ and threshold parameters $s_0$. On the other hand, the dominant contributions come from the perturbative terms, the operator product  expansion is well convergent.

\begin{figure}
 \centering
 \includegraphics[totalheight=5cm,width=6cm]{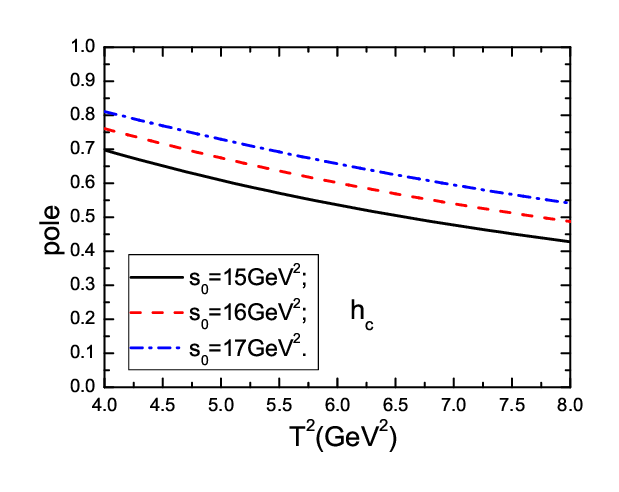}
\includegraphics[totalheight=5cm,width=6cm]{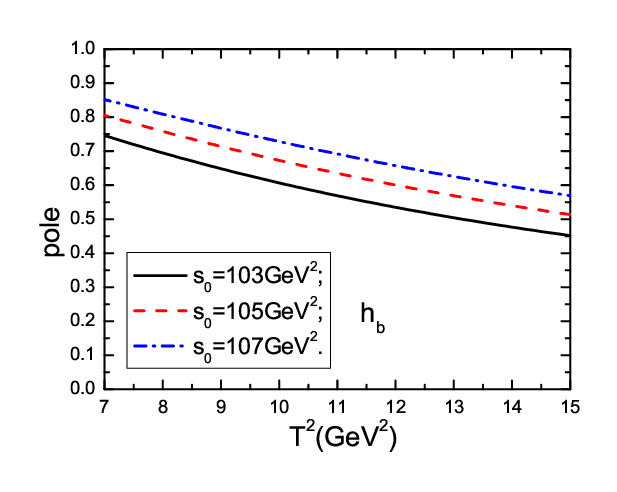}
        \caption{ The pole contributions with  variations of the Borel parameters $T^2$.}
\end{figure}

 In Fig.12, we plot the  $h_Q$ masses $M_{h_Q}$ and decay constants $f_{h_Q}$ with  variations of the Borel parameters $T^2$ and threshold parameters $s_0$.
 From the figure, we can see that the values are stable with variations of the Borel parameters $T^2$. In Fig.13, we plot the  masses  $M_{h_Q}$ and decay constants $f_{h_Q}$ with  variations of the moment parameters $n$ and threshold parameters $s_0$ in the moments sum rules. In the moments sum rules for the $P$-wave heavy quarkonium states, $\xi>1$ \cite{Reinders85}, in this article, we take $\xi=2$, and choose $n=3-7$ and $n=17-23$ for the $h_c$ and $h_b$, respectively. From  Figs.12-13, we can see that the values from the moments sum rules are consistent with that from the Borel sum rules.

\begin{figure}
 \centering
 \includegraphics[totalheight=5cm,width=6cm]{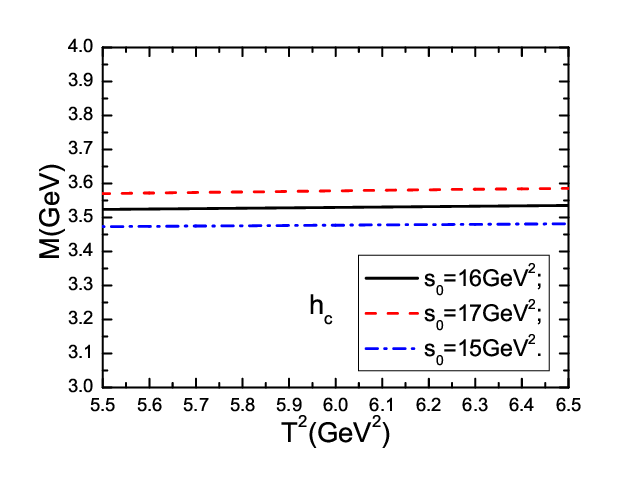}
\includegraphics[totalheight=5cm,width=6cm]{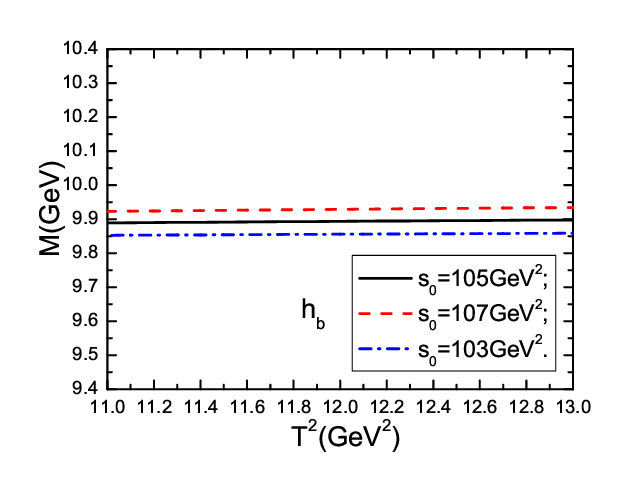}
\includegraphics[totalheight=5cm,width=6cm]{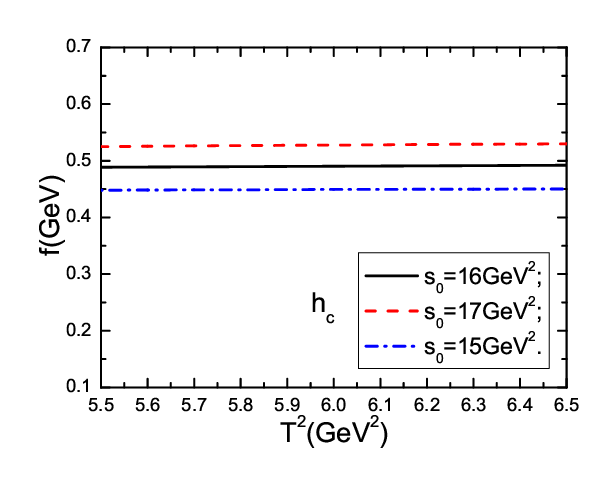}
\includegraphics[totalheight=5cm,width=6cm]{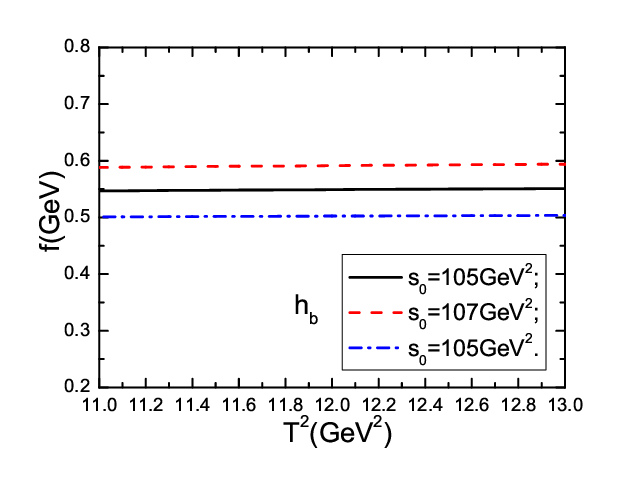}
        \caption{ The masses  $M_{h_Q}$ and decay constants $f_{h_Q}$ with  variations of the Borel parameters $T^2$.}
\end{figure}

 In the following, we write down  the masses and decay constants of
 the  heavy quarkonium states $h_c$ and $h_b$,
\begin{eqnarray}
M_{h_c}&=&3.530\pm0.006\pm0.050 \pm0.090\,\rm{GeV} \, ,  \nonumber\\
M_{h_b}&=&9.894\pm 0.005\pm0.035\pm0.064\,\rm{GeV} \, ,  \nonumber\\
f_{h_c}&=&0.490\pm0.002\pm0.040\pm0.045\,\rm{GeV} \, ,  \nonumber\\
f_{h_b}&=&0.549\pm0.002\pm0.050\pm0.045\,\rm{GeV} \, ,
\end{eqnarray}
from the Borel sum rules, and
\begin{eqnarray}
M_{h_c}&=&3.521\pm0.025\pm0.050\pm0.098\,\rm{GeV} \, ,  \nonumber \\
M_{h_b}&=&9.899\pm0.006\pm0.040\pm0.063\,\rm{GeV} \, ,  \nonumber \\
f_{h_c}&=&0.490\pm0.008\pm0.040\pm0.044\,\rm{GeV} \, ,   \nonumber\\
f_{h_b}&=&0.552\pm0.003\pm0.047\pm0.046\,\rm{GeV} \, ,
\end{eqnarray}
from the moments sum rules. The uncertainties come from the Borel parameters (or moment parameters), threshold parameters, heavy quark masses, sequentially.
    The integral ranges $4m_Q^2\sim s_0$ and the QCD spectral densities change quickly with variations of the heavy quark masses, small variations $\delta m_Q$ can lead to relatively large uncertainties $\delta M_{h_Q}$ and $\delta f_{h_Q}$.
  In this article, we take $\delta 4m_c^2=8m_c\delta m_c=\pm 1 \,\rm{GeV}^2$ and $\delta 4 m_b^2=8m_b\delta m_b=\pm 2 \,\rm{GeV}^2$, just like the uncertainties of the continuum  threshold parameters $\delta s^0_{h_c}=\pm 1\,\rm{GeV}^2$ and $\delta s^0_{h_b}=\pm 2\,\rm{GeV}^2$.
The masses from both QCD sum rules are consistent  with the experimental data, $M_{h_b(1\rm{P})}=\left(9898.3\pm1.1{}^{+1.0}_{-1.1}\right)\, \rm{MeV}$ \cite{Belle1103} and $M_{h_c(1\rm{P})}=\left(3525.41\pm0.16\right)\,\rm{MeV}$ \cite{PDG}.
  The heavy quarkonium states $h_Q$ couple potentially to the tensor currents $\bar{Q}\sigma_{\mu\nu}Q$, the $h_Q$ have the quark structure $\epsilon^{ijk}\xi^{\dagger}\sigma^k\zeta$ besides the quark structure $ik_2^i \xi^{\dagger}\sigma \cdot (\vec{k}_1-\vec{k}_2)\zeta$.

\begin{figure}
 \centering
 \includegraphics[totalheight=5cm,width=6cm]{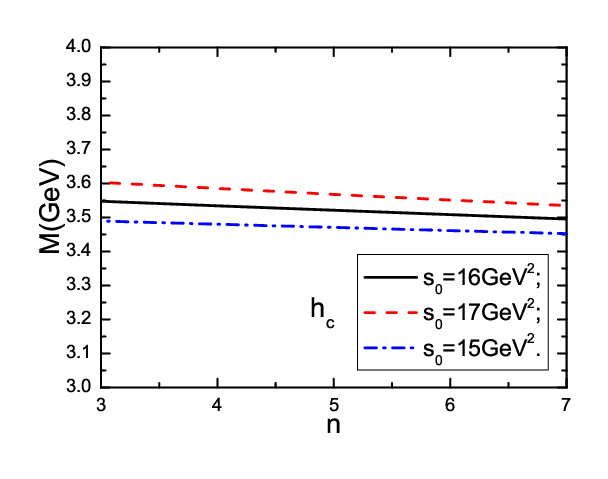}
\includegraphics[totalheight=5cm,width=6cm]{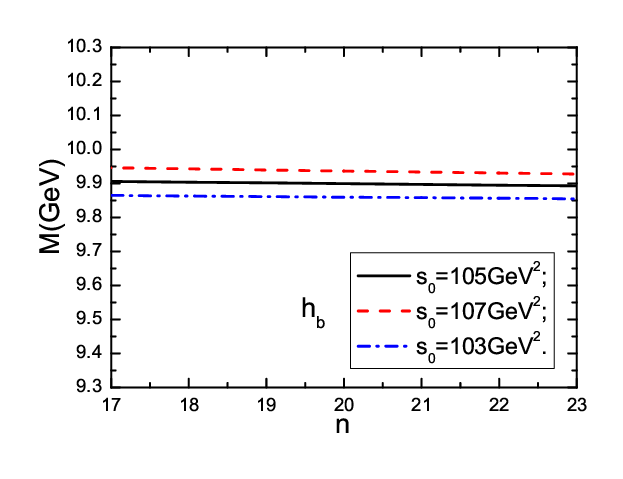}
\includegraphics[totalheight=5cm,width=6cm]{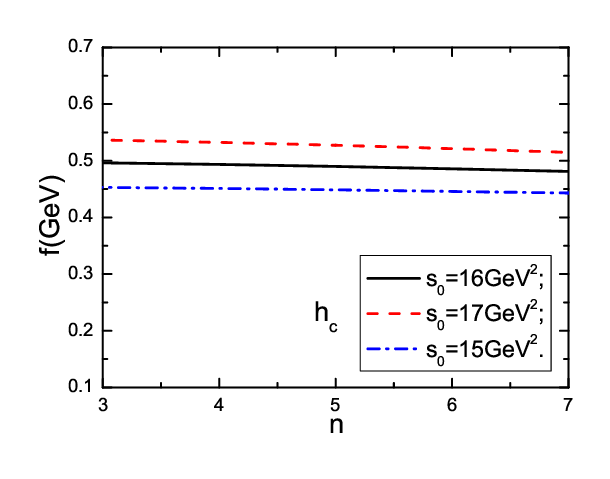}
\includegraphics[totalheight=5cm,width=6cm]{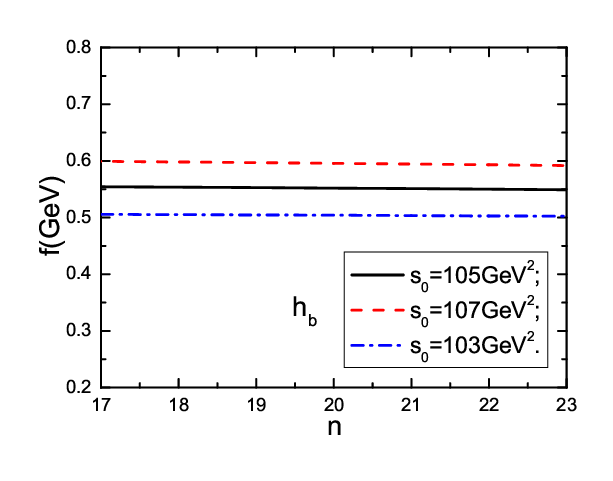}
        \caption{ The masses  $M_{h_Q}$ and decay constants $f_{h_Q}$ with  variations of the  moment parameters $n$.}
\end{figure}

For the heavy quarkonium states, especially for the bottomonium states, the relative velocities $\omega$ of the quarks  are small, we should account for the Coulomb-like $\frac{\alpha_s^\mathcal{C}}{\omega}$  corrections.  After taking into account  all the Coulomb-like contributions, we obtain the coefficient $f(\omega)$ to dress   the leading-order spectral densities  $\rho_0(s)$ \cite{Coulomb-BC},
\begin{eqnarray}
f(\omega)&=&\frac{4\pi\alpha_s^\mathcal{C}}{3\omega} \frac{1}{1-\exp\left(-\frac{4\pi\alpha_s^\mathcal{C}}{3\omega}\right)}=1+\frac{2\pi\alpha_s^\mathcal{C}}{3\omega}+\cdots\,   .
\end{eqnarray}
In Fig.14, we plot the coefficients $f(\omega)=1+\frac{\rho_1(s)}{\rho_0(s)}$ and $1+\frac{2\pi\alpha_s}{3\omega}$ for the heavy quarkonium states $h_c$ and $h_b$, respectively, and take the approximation $\alpha_s^{\mathcal{C}}=\alpha_s$. From the
figure, we can see that $1+\frac{\rho_1(s)}{\rho_0(s)}\approx1+\frac{2\pi\alpha_s}{3\omega}$, the perturbative $\alpha_s$ corrections $\rho_1(s)$ can be  approximated  by $\rho_0(s)\frac{2\pi\alpha_s}{3\omega}$. We can   account for all the   Coulomb-like contributions by multiplying the leading-order spectral densities $\rho_0(s)$ by the coefficient $f(\omega)$ tentatively. If we take the Borel parameters as $T^2=(6.8-7.8)\,\rm{GeV}^2$ and $(12.9-14.9)\,\rm{GeV}^2$ in the channels $h_c$ and $h_b$, respectively,   again we obtain the pole contributions $(51-67)\%$ and $(50-67)\%$, respectively.
The central values
\begin{eqnarray}
M_{h_c}&=&3.516 \,\rm{GeV} \, ,  \nonumber\\
M_{h_b}&=&9.884 \,\rm{GeV} \, ,  \nonumber\\
f_{h_c}&=&0.576\,\rm{GeV} \, ,  \nonumber\\
f_{h_b}&=&0.657\,\rm{GeV} \, ,
\end{eqnarray}
come from the Borel sum rules indicate the shifts  $\delta M_{h_c}=-0.014\,\rm{GeV}$, $\delta M_{h_b}=-0.010\,\rm{GeV}$, $\delta f_{h_c}=0.086\,\rm{GeV}$, $\delta f_{h_b}=0.108\,\rm{GeV}$ compared to the predictions  in Eq.(35). The mass-shifts are mild, while the decay constant shifts are large.

\begin{figure}
 \centering
 \includegraphics[totalheight=6cm,width=8cm]{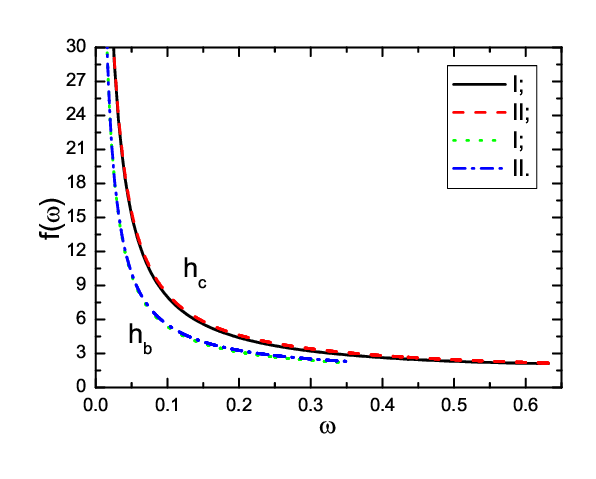}
    \caption{ The  coefficients   $f(\omega)=1+\frac{\rho_1(s)}{\rho_0(s)}$ and $1+\frac{2\pi\alpha_s}{3\omega}$ for I and II, respectively. }
\end{figure}

In the $q\bar{q}$ quark model, the party $P=(-1)^{L+1}$, the charge conjunction $C=(-1)^{L+S}$, where $L$ and $S$ are the orbital and spin angular momenta, respectively. The heavy quarkonium states $h_Q$ have $J^{PC}=1^{+-}$, so they have the quantum numbers  $S=0$, $L=1$ and $J=L$, the spins of the quark $Q$ and antiquark $\bar{Q}$   should be antiparallel. The quark structures $\epsilon^{ijk}\xi^{\dagger}\sigma^k\zeta$ and $ik_2^i \xi^{\dagger}\sigma \cdot (\vec{k}_1-\vec{k}_2)\zeta$ both satisfy the requirement, the heavy quarkonium states $h_Q$ have two possible quark structures. We can study the mixing of the two structures with the two-point correlation functions
$\Pi_{\mu\alpha}(p)$,
\begin{eqnarray}
\Pi_{\mu\alpha}(p)&=&i\int d^4x e^{ip \cdot x} \langle
0|T\left\{\eta_{\mu}(x)\eta_{\alpha}^{\dagger}(0)\right\}|0\rangle \, , \nonumber\\
\eta_{\mu}(x)&=&\cos\theta \, \bar{Q}(x)\sigma_{\mu\nu}\gamma_5 Q(x)p^\nu+\sin\theta\,\bar{Q}(x)\partial_{\mu }\gamma_5 Q(x)\, ,
\end{eqnarray}
and search for the optimal value of the mixing angular $\theta$.

\section{Conclusion}
In this article, we take the tensor currents $\bar{Q}(x)\sigma_{\mu\nu}Q(x)$ to interpolate the $P$-wave spin-singlet heavy quarkonium states $h_Q$,  study the masses and decay constants with  the Borel sum rules and the moments sum rules, and explore whether or not the $h_Q$ have the quark structure $\epsilon^{ijk}\xi^{\dagger}\sigma^k\zeta$ besides the quark structure $ik_2^i \xi^{\dagger}\sigma \cdot (\vec{k}_1-\vec{k}_2)\zeta$.
The  masses and decay constants come from the Borel sum rules and moments sum rules are consistent with each other, the masses are also consistent with the experimental data. The heavy quarkonium states $h_Q$ couple potentially to the tensor currents $\bar{Q}(x)\sigma_{\mu\nu}Q(x)$, and have the quark structure $\epsilon^{ijk}\xi^{\dagger}\sigma^k\zeta$ besides the quark structure $ik_2^i \xi^{\dagger}\sigma \cdot (\vec{k}_1-\vec{k}_2)\zeta $.
We can take the decay constants as   basic input parameters and study the revelent hadronic
processes with the QCD sum rules, for example, we can study the $h_c DD^*$, $ h_c D_sD_s^*$, $h_c D^*D^*$, $h_c D_s^*D_s^*$ vertexes and  $h_c \to D,\, D_s,\,D^*,\,D_s^*$ form-factors with three-point correlation functions.
In calculations, we take into account the leading-order, next-to-leading-order perturbative contributions, and the gluon condensate, four-quark condensate contributions in the operator product expansion.
 The analytical expressions of the perturbative spectral densities have applications in studying the two-body decays of a boson to two fermions with the vertexes $\sigma_{\mu\nu}\gamma_5$ and $\sigma_{\mu\nu}$.

\section*{Acknowledgements}
This  work is supported by National Natural Science Foundation,
Grant Number 11075053,  and the Fundamental Research Funds for the
Central Universities.

\section*{Appendix}
We take the notation
\begin{eqnarray}
\int d ps&=& \int \frac{d^{D-1}\vec{k}}{2E_k}\frac{d^{D-1}\vec{p}_1}{2E_{p_1}}\frac{d^{D-1}\vec{p}_2}{2E_{p_2}}\delta^D(p-k-p_1-p_2) \, ,\nonumber
\end{eqnarray}
for simplicity, and write down the analytical expressions of the three-body phase-space integrals,
 \begin{eqnarray}
R_{12}(s)&=& \frac{s  }{\pi^2\sqrt{\lambda(s,m_Q^2,m_Q^2)}}(2\pi)^{-4\varepsilon_{\rm IR}}\mu^{-2\varepsilon_{\rm IR}}\int d ps \frac{1}{k\cdot p_1 k\cdot p_2 }\nonumber\\
&=&\frac{1}{\sqrt{\lambda(s,m_Q^2,m_Q^2)}} \left\{\log\left(\frac{1+\omega}{1-\omega}\right)\left[ \frac{1}{\varepsilon_{\rm IR}}-2\log4\pi+2\gamma-2+2\log\frac{\sqrt{\lambda(s,m_Q^2,m_Q^2)}^3}{m_Q^2s\mu^2}\right] \right. \nonumber \\
&&\left.-\log^2\left(\frac{1+\omega}{1-\omega}\right)-4{\rm Li_2}\left( \frac{2\omega}{1+\omega}\right) -{\rm Li_2}\left( \frac{1+\omega}{2}\right)-2{\rm Li_2}\left( \omega\right)+\log2\log(1+\omega) -\frac{\log^2 2}{2}+\frac{\pi^2}{12}\right\}\, ,\nonumber\\
&=&\overline{R}_{12}(s)+\frac{1}{\sqrt{\lambda(s,m_Q^2,m_Q^2)}} \log\left(\frac{1+\omega}{1-\omega}\right)\left[ \frac{1}{\varepsilon_{\rm IR}}-2\log4\pi+2\gamma-2+2\log\frac{\sqrt{\lambda(s,m_Q^2,m_Q^2)}^3}{m_Q^2s\mu^2}\right] \, , \nonumber
\end{eqnarray}
\begin{eqnarray}
R^1_{12}(s)&=& \frac{s  }{\pi^2\sqrt{\lambda(s,m_Q^2,m_Q^2)}}\int d ps \frac{s-K^2}{k\cdot p_1 k\cdot p_2 }\nonumber\\
&=&\frac{s}{\sqrt{\lambda(s,m_Q^2,m_Q^2)}} \left\{\log^2(1-\omega)-\log^2(1+\omega) +2\log2\log\left(\frac{1+\omega}{1-\omega}\right)
+2{\rm Li_2}\left(\frac{1-\omega}{2}\right)\right. \nonumber\\
&&\left.-2{\rm Li_2}\left(\frac{1+\omega}{2}\right) \right\} \, ,\nonumber \\
R^2_{12}(s)&=& \frac{s}{\pi^2\sqrt{\lambda(s,m_Q^2,m_Q^2)}}\int d ps \frac{\left(s-K^2\right)^2}{k\cdot p_1 k\cdot p_2 }\nonumber\\
&=&\frac{s^2}{\sqrt{\lambda(s,m_Q^2,m_Q^2)}} \left\{\log^2(1-\omega)-\log^2(1+\omega) +2\log4\log\left(\frac{1+\omega}{1-\omega}\right)
+2{\rm Li_2}\left(\frac{1-\omega}{2}\right)\right. \nonumber \\
&&\left.-2{\rm Li_2}\left(\frac{1+\omega}{2}\right)
 +2\omega  -(1+\omega^2)\log\left(\frac{1+\omega}{1-\omega}\right) \right\} \, , \nonumber
\end{eqnarray}
\begin{eqnarray}
R^1_{0}(s)&=& \frac{1  }{\pi^2}\int d ps  (s-K^2) \nonumber\\
&=&\frac{s\sqrt{\lambda(s,m_Q^2,m_Q^2)}}{96} \left\{\omega(15-4\omega^2-3\omega^4) +\frac{3}{2}(\omega^6+\omega^4+3\omega^2-5)\log\left(\frac{1+\omega}{1-\omega}\right)
 \right\} \, .\nonumber\\
\end{eqnarray}

\end{document}